\pdfoutput=1
\documentclass{article}
\usepackage{cite,spconf,amsmath,graphicx,enumitem,amssymb,color,bm,subfigure}
\def\x{{\bm{x}}}
\def\y{{\bm{y}}}
\def\s{{\bm{s}}}

\def\n{{\bm{n}}}
\def\fx{{f_{\theta_x}}}
\def\fy{{f_{\theta_y}}}
\def\g{{g_{\phi}}}
\def\F{\bm{F}}

\title{Distributed Image Transmission using Deep Joint Source-Channel Coding}

\name{Sixian~Wang $^{\star}$, Ke~Yang$^{\star}$, Jincheng~Dai$^{\star}$, Kai~Niu$^{\star \dagger}$
\thanks{This work was supported in part by the National Natural Science Foundation of China under Grant 92067202, Grant 62071058, and Grant 62001049, the Beijing Municipal Natural Science Foundation under Grant 4222012, and the Fundamental Research Funds for the Central Universities. Corresponding author: Jincheng Dai.}
}

\address{
	\normalsize $^{\star}$ Beijing University of Posts and Telecommunications, Beijing, 100876, China \\
	\normalsize $^{\dagger}$ Peng Cheng Laboratory, Shenzhen, China
}

\begin{document}
\ninept
\maketitle
 
\begin{abstract}
We study the problem of deep joint source-channel coding (D-JSCC) for correlated image sources, where each source is transmitted through a noisy independent channel to the common receiver. In particular, we consider a pair of images captured by two cameras with probably overlapping fields of view transmitted over wireless channels and reconstructed in the center node. The challenging problem involves designing a practical code to utilize both source and channel correlations to improve transmission efficiency without additional transmission overhead. 
To tackle this, we need to consider the common information across two stereo images as well as the differences between two transmission channels.
In this case, we propose a deep neural networks solution that includes lightweight edge encoders and a powerful center decoder. Besides, in the decoder, we propose a novel channel state information aware cross attention module to highlight the overlapping fields and leverage the relevance between two noisy feature maps.Our results show the impressive improvement of reconstruction quality in both links by exploiting the noisy representations of the other link. 
Moreover, the proposed scheme shows competitive results compared to the separated schemes with capacity-achieving channel codes.
\end{abstract}


\section{Introduction}
With the rapid development of mobile terminals and wireless networks, applications of multi-node communications like multi-camera surveillance systems, IoT networks, 3D scene capture, and stereo image transmission has drawn a lot of attention. 
The main challenge is the asymmetric distribution of computational capability and power budget between edge devices and center sever, making the naive use of conventional separated communication systems unsuitable.
A promising solution is the distributed source coding (DSC), which exploits the statistics of distributed sources only at the receiver side, enabling low-complexity encoding by shifting the bulk of computation to the receiver. Based on the information-theoretic bounds given by Slepian and Wolf \cite{SW} for distributed lossless coding, Wyner and Ziv extended it to lossy compression with side information at the decoder \cite{WZ}. After that, practical Slepian-Wolf and Wyner-Ziv schemes have become research hot pots. However, most distributed source coding techniques are derived from proven channel coding ideas and concentrate on synthetic datasets, and specific correlation structures \cite{DSC1,DSC2}. 

Thanks to the boosting development of deep learning (DL), recent advances achieve practical lossy compression schemes \cite{DSIN,DWSIC} and JSCC schemes \cite{xuan2021low} for high-dimensional image sources using deep neural networks (DNNs), which show substantial improvements by exploiting decoder side information. In this paper, taking this idea one step further, we consider the design of practical distributed joint source-channel coding (JSCC) scheme. 
Although it has been proven that the separation theorem still holds in distributed scenarios, limited by delay and complexity in the distributed sceneries, distributed JSCC can reduce signal distortion and obtain a better performance through the integration of source and channel codes. Another relevant work is the deep JSCC (D-JSCC) schemes \cite{DJSCC, DJSCCF, ADJSCC, wang2021novel}, which achieves graceful degradation as the channel signal-to-noise ratio (SNR) varies and outperforms the state-of-the-art digital schemes. In the distributed scenario, the scheme needs to be redesigned.

In this paper, we propose a novel distributed D-JSCC scheme for wireless image transmission. The proposed scheme considers the \emph{independent channels case} \cite{DSC3}, where each source is transmitted through a noisy independent channel and recovered jointly at the common receiver. 
In particular, each pair of correlated images come from two cameras with probably overlapping fields of view. The proposed distributed encoders and the decoder are trained jointly to exploit such common dependence. 
Meanwhile, we propose a channel state information-aware cross attention module to enable efficient information interaction between two noisy feature maps at the receiver. It measures the correlation between the images in patch level, where the channel quality and the spatial correlation are explicitly considered. In this manner, the decoder can better reconstruct both images without extra transmission overhead to improve the whole transmission efficiency.  We compare the proposed scheme to various image codecs combined with ideal capacity-achieving channel code or practical LDPC codes. The considered image coding algorithms include popular image codecs - BPG and JPEG2000 and the recently developed DL-based DSC image compression scheme. Results demonstrate we achieve an apparent performance gain compared to the baseline scheme without jointly decoding, and the improvement depends on the difference of SNRs between the two links. Besides, our method shows impressive performance on PSNR and MS-SSIM metrics compared to all schemes. 

\section{System Model and Proposed Architecture}
\begin{figure*}[t]
	\centering
	\includegraphics[scale=0.45]{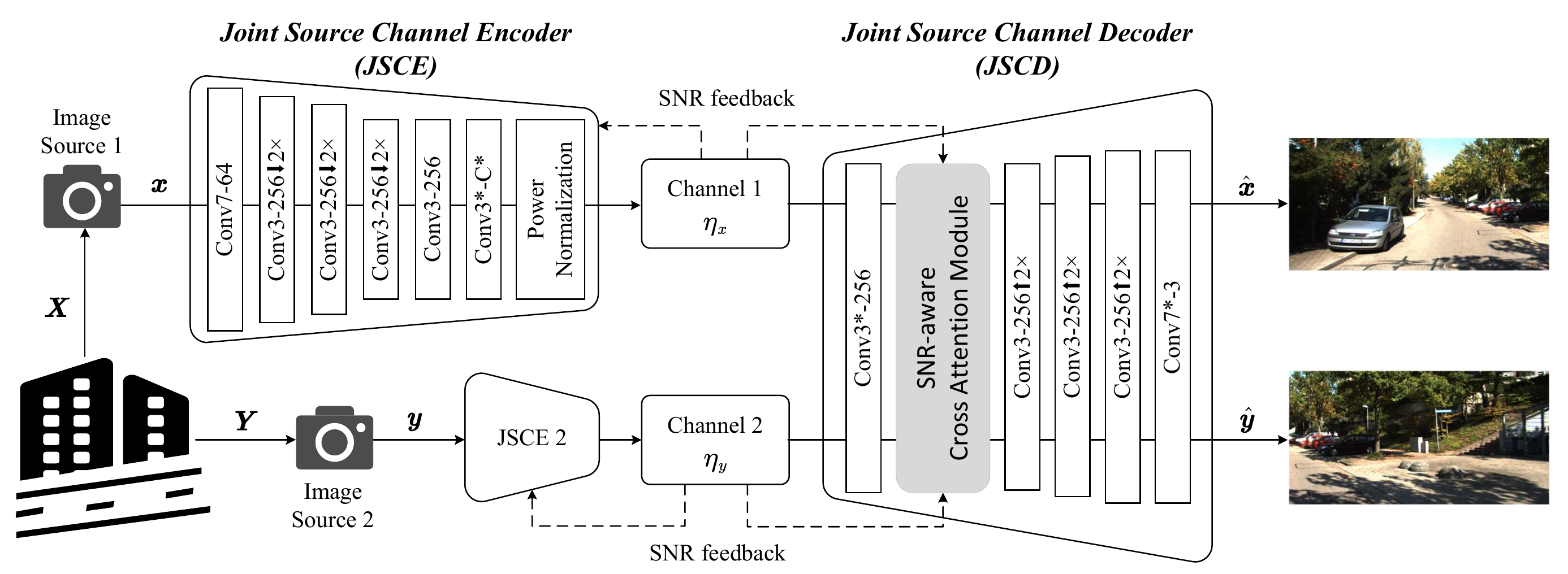}
	\caption{System model of the independent channels case and the proposed encoder and decoder structure.}  
	\vspace{-0.4cm}
	\label{SystemArchtecture}
\end{figure*}
Fig. \ref{SystemArchtecture} shows the system model, consider two statistically dependent identically distributed (i.i.d.) image sources $\bm{X}$ and $\bm{Y}$, let $\x$ and $\y$ represent samples of $\bm{X}$ and $\bm{Y}$ from possible infinite sets $\mathcal{X}$ and $\mathcal{Y}$. 
The joint source channel encoder (JSCE) is a deterministic encoding function denoted as $\fx$, which firstly maps an $n$-dimensional input image $\bm{x} \in \mathbb{R}^n$ into complex symbols $\bm{s}_x \in \mathbb{C}^{k}$, and employs a power normalization module to ensure the average power constraint, i.e., $\s_x$ is normalized as $\tilde{\s}_x = \sqrt{kP} \frac{\s_x}{\left\| \s_x \right\|_2}$ before transmission. For the additive white Gaussian noise (AWGN) channels, the transfer function is $\eta_x (\tilde{\s}_x) = \tilde{\s}_x + \n_x$, where $\n \in \mathbb{C}^{k}$ is the i.i.d. Gaussian noise with zero mean and variance $\sigma_x^2$.  
We assume the SNR defined as $\mu_x = 10\log_{10}{\frac{P}{\sigma_x^2}}$ of every link can be different and known at transmitter and receiver.
In the considered scenario, the other input $\bm{y} \in \mathbb{R}^n$ goes through the same process, and the two JSCEs are working independently, i.e., one image is encoded without access to the other. 

The joint source-channel decoder (JSCD) denoted as $\g$ decodes two noisy received symbols $\hat{\s}_x$ and $\hat{\s}_y$, and produce their reconstructions as 
\begin{equation}
\hat{\x} = \g\left( \eta_x\left( \fx\left( \x,\mu_x \right) \right), \eta_y\left( \fy\left( \y,\mu_y\right) \right), \mu_x \right),
\end{equation}
\begin{equation}
\hat{\y} = \g\left( \eta_y\left( \fy\left( \y,\mu_y \right) \right), \eta_x\left( \fx\left( \x,\mu_x \right) \right),  \mu_y \right),
\end{equation}
where $\theta_x$ and $\theta_y$ are the parameters set of JSCEs, and $\phi$ is the shared JSCD's. 
Although it seems to be a viable alternative that recovers the image from a high-quality link first and then uses it as side information to help decode the other. 
In practice, the channel quality of each node may vary a lot with time, and mismatched or low-quality side information cause degradation in the reconstruction quality. Besides, the serial decoding manner increases transmission latency and requires training and storage of multiple decoders, which is unfriendly to practical distributed applications.
As a consequence, we propose a parallel decoding framework, which only uses one group of parameters to recover all images at the same time. 
To be more specific, the JSCD jointly takes two noisy feature maps (received symbol vectors) as well as their SNRs as input. The recovery process is fully parallel by concatenating two feature maps in the batch dimension. 

In addition, the proposed architecture can also adapt to the \emph{asymmetric case} \cite{DSC3}, which assumes one of the sources is lossless available at the receiver and used as side information during the decoding of the other source. We also train and report this case's performance by setting the SNR of the available source to $+\infty$. The system of every case is trained to minimize the empirical distortion loss defined as 
\begin{equation}
\mathcal{L} = \mathbb{E}_{\bm{X},\bm{Y}}\left[ d\left( \x, \hat{\x} \right) + \alpha d\left( \y, \hat{\y} \right) \right] 
\end{equation} 
where $d(\cdot,\cdot)$ is the distortion between the input image and the reconstruction, and hyperparameter $\alpha$ determines the relative importance of two images. 

In the proposed architecture, both JSCE and JSCD are parameterized by a series of convolutional blocks, and we plot the detailed structure in Fig. \ref{SystemArchtecture}.  Conv$K$-$C_\text{out}$ denotes a convolutional block including a convolutional layer with $K \times K$ kernel size and $C_\text{out}$ filters, a ChannelNorm layer \cite{HIFIC}, ReLU activation function, and a attention feature module in turn. ``$\uparrow$'' / ``$\downarrow$'' mark the upscaling/downscaling convolutions with stride $S=2$, and ``*'' indicates a single convolution used for compressing/decompressing the feature map into/from $C^{*}$ channels to satisfy the given bandwidth rate. Besides, we employ the attention feature module proposed in \cite{ADJSCC} to reduce the training cost, which can enable a single model to deal with different SNRs.
\section{SNR-aware Cross Attention Mechanism \\ at the receiver}

The existing methods in DNN based DSC achieve efficient lossy compression by using a correlated lossless image as the side information. 
The considered \emph{independent channels case} is a more realistic scenario, where the mutual information changes with each pair of samples and both channel conditions. The lossy or uncorrelated side information may cause the decrease of recovered image quality. Moreover, we can no longer assume that the two images are always aligned nor that the layout of the images is similar. Thus, for different images and different channel conditions, it requires a dynamic decision algorithm to measure and make use of the correlation. 
To conquer this challenge, we propose an SNR-aware cross attention module (SCAM) to achieve beneficial information interactions between two samples using noisy feature maps, where the feature map of one image is dynamically adjusted according to the other. Meanwhile, the channel quality and the spatial correlation are jointly considered. 

\begin{figure}[t]
	\centering
	\includegraphics[scale=0.4]{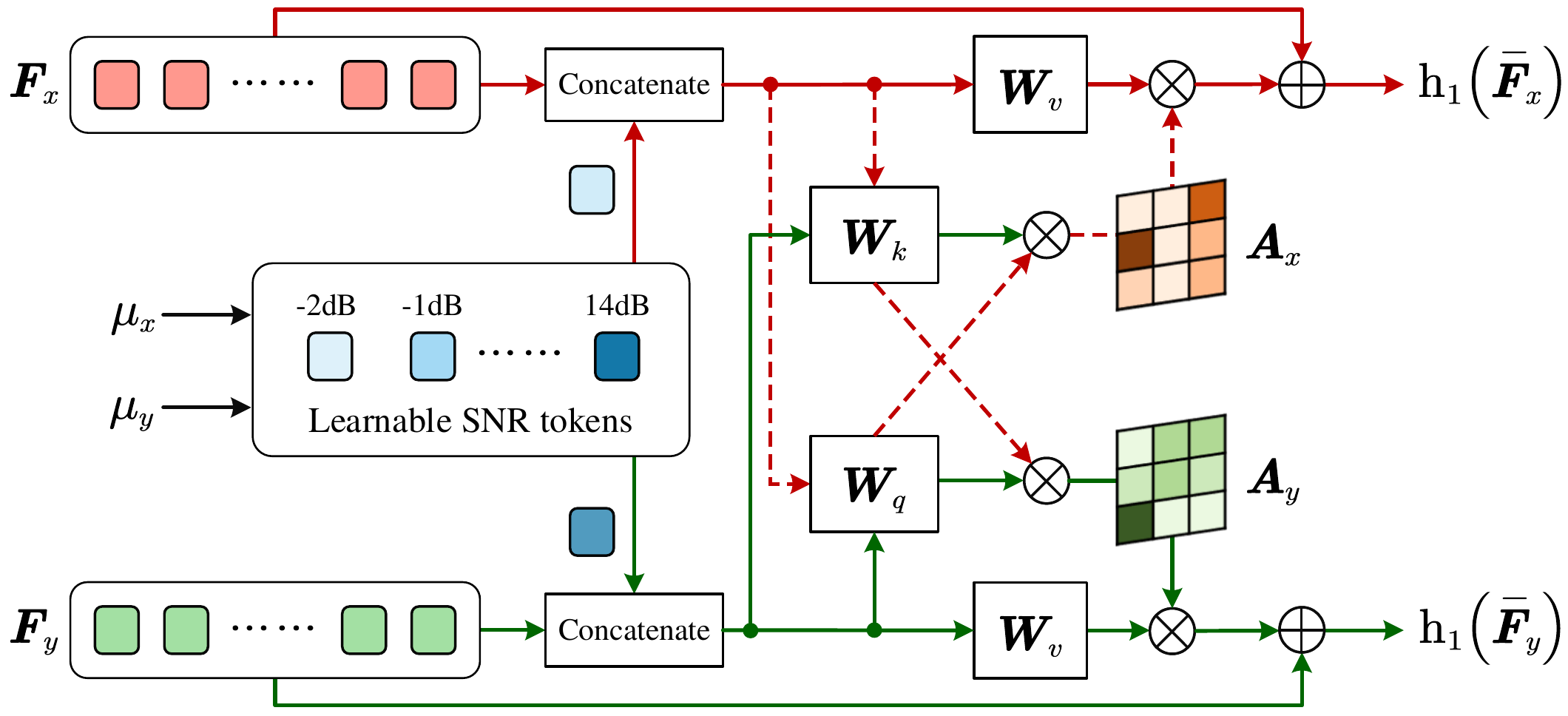}
	\caption{The details of the proposed SNR-aware Cross Attention Module. For a better visualization, each square indicates a slice of $\bm{F}$ in spatial position, which is a vector in $\mathbb{R}^{1 \times C}$.}  
	\vspace{-0.4cm}
	\label{CAM}
\end{figure}

Inspired by vision transformers \cite{VIT}, which uses a self-attention mechanism to capture connections and dependencies between global and local content, the proposed SCAM is shown in Fig. \ref{CAM}. For simplicity, we use $x$ and $y$ to denote the stereo images and omit subscripts when they need not be distinguished. The two noisy feature maps $\F_x \in \mathbb{R}^{C \times H_x \times W_x}$ and $\F_y \in \mathbb{R}^{C \times H_y \times W_y}$ is extracted from the previous convolution block, where $H$, $W$, and $C$ denote the height, width, and the number of the channel of the feature maps respectively. SCAM aims to generate cross attention maps, which are further used to recalibrate the feature maps to achieve information interactions between two feature maps. We use patches to measure the correction between images, the feature map $\bm{F} \in \mathbb{R}^{C\times H \times W} $ is reshaped to $\mathbb{R}^{M \times C}$, where $M = H \times W$ can be viewed as the number of spatial dimensions. Note that, since the encoder uses $2\times$ downscaling for three times, a vector here in spatial dimension is extracted from a patch of $8 \times 8$ pixels in the same position. 
It consists of the following steps in turn: \\
\textbf{SNR Information Attachment.} To adapt to a range of SNRs and promise considerable performance gain when the SNR of two links is different, it is important to let channel state information interact with context information. Thus, we inform the network of current channel quality by pre-setting a group of learnable quality tokens $\mathcal{S}=\left\lbrace \bm{S}_1,\bm{S}_2,\cdots,\bm{S}_m \right\rbrace$. Each token $\bm{S}_i \in \mathbb{R}^{1 \times C}$ is used to cover a range of SNR values. If SNR of the current transmission channel is covered by $\bm{S}_j$, we concatenate it with $\F_x$($\F_y$) in the spatial dimension:
\begin{equation}
\overline{\bm{F}} = \left[ \bm{S}_j, \bm{F}^{(1)}, \bm{F}^{(2)}, \cdots ,\bm{F}^{(M)} \right]
\end{equation}
where $\overline{\bm{F}} \in \mathbb{R}^{(M+1) \times C }$ is the fused feature map including channel quality and context information.\\
\textbf{Cross Attention Layer.} To evaluate the correlation of two sources, we employ three linear layers $\bm{W}_q,\bm{W}_k,\bm{W}_v \in \mathbb{R}^{C \times C}$ to map the input to its query $\bm{Q}$, key $\bm{K}$, and value $\bm{V}$, \textit{i.e.}, $\bm{Q}=\varphi \left( \overline{\bm{F}} \right) \bm{W}_q$, $\bm{K} = \varphi \left( \overline{\bm{F}} \right) \bm{W}_k$, and $\bm{V} = \varphi \left( \overline{\bm{F}} \right) \bm{W}_v$, where $\varphi$ refers to layer normalization function. The cross attention is calculated between spatial vectors of two feature maps, that is, let query vectors of $\overline{\F_x}$ ($\overline{\F_y}$) to multiply the key value of $\overline{\F_y}$ ($\overline{\F_x}$) to get cross attention map $\bm{A}_x = \bm{Q}_x  \bm{K}_y^\mathrm{T}$ ($\bm{A}_y = \bm{Q}_y  \bm{K}_x^\mathrm{T}$). In this manner, the relevance between $\overline{\bm{F}}_x^{(i)}$ and $\overline{\bm{F}}_y^{(i)}$ is calculated, which explicitly concerns the interaction between channel state information and context information. 
\textbf{Feature map recalibration.}
We use cross attention map to recalibrate the value matrix:
\begin{equation}
h_1\left( \overline{\bm{F}}\right)  = 
\overline{\bm{F}} + \sigma_1\left( \frac{ \bm{A}}{\sqrt{C}} \right) \bm{V} \bm{W}_o 
\end{equation}  
where $\bm{W}_o \in \mathbb{R}^{C \times C}$ is a linear layer, $\sigma_1$ represent the SoftMax operator applied to each column of the matrix for normalization. After that, to further fuse the context and channel state information, we adapt a multilayer perceptron (MLP) including two linear layers with a skip connection:
\begin{equation}
\begin{aligned}
h_2\left( \overline{\bm{F}} \right) = h_1\left( \overline{\bm{F}} \right) + \sigma_2 \left( 
\varphi \left(h_1\left( \overline{\bm{F}} \right)\right)   
\bm{W}_1 + \bm{b}_1 \bm{1}_C^\mathrm{T}
\right) \bm{W}_2 + \bm{b}_2 \bm{1}_C^\mathrm{T}
\end{aligned}
\end{equation}
where $\bm{W}_1 \in \mathbb{R}^{C \times C_h}$, $\bm{W}_2 \in \mathbb{R}^{C_h \times C}$ are linear layers, $\bm{b}_1 \in \mathbb{R}^{C_h}$,$\bm{b}_2 \in \mathbb{R}^{C}$ are bias, $\sigma_2$ is the ReLU activation function, and $C_h$ is the hidden layer size of the MLP.
In the end of cross attention layer, we remove the attached quality tokens from $h_2\left( \overline{\bm{F}}\right)$ and finally get the recalibrated feature map $\tilde{\bm{F}}$.

\section{Experiments} 
\begin{figure*}[htbp]
	\begin{center}
		\subfigure[]{
			\includegraphics[width=0.24\textwidth]{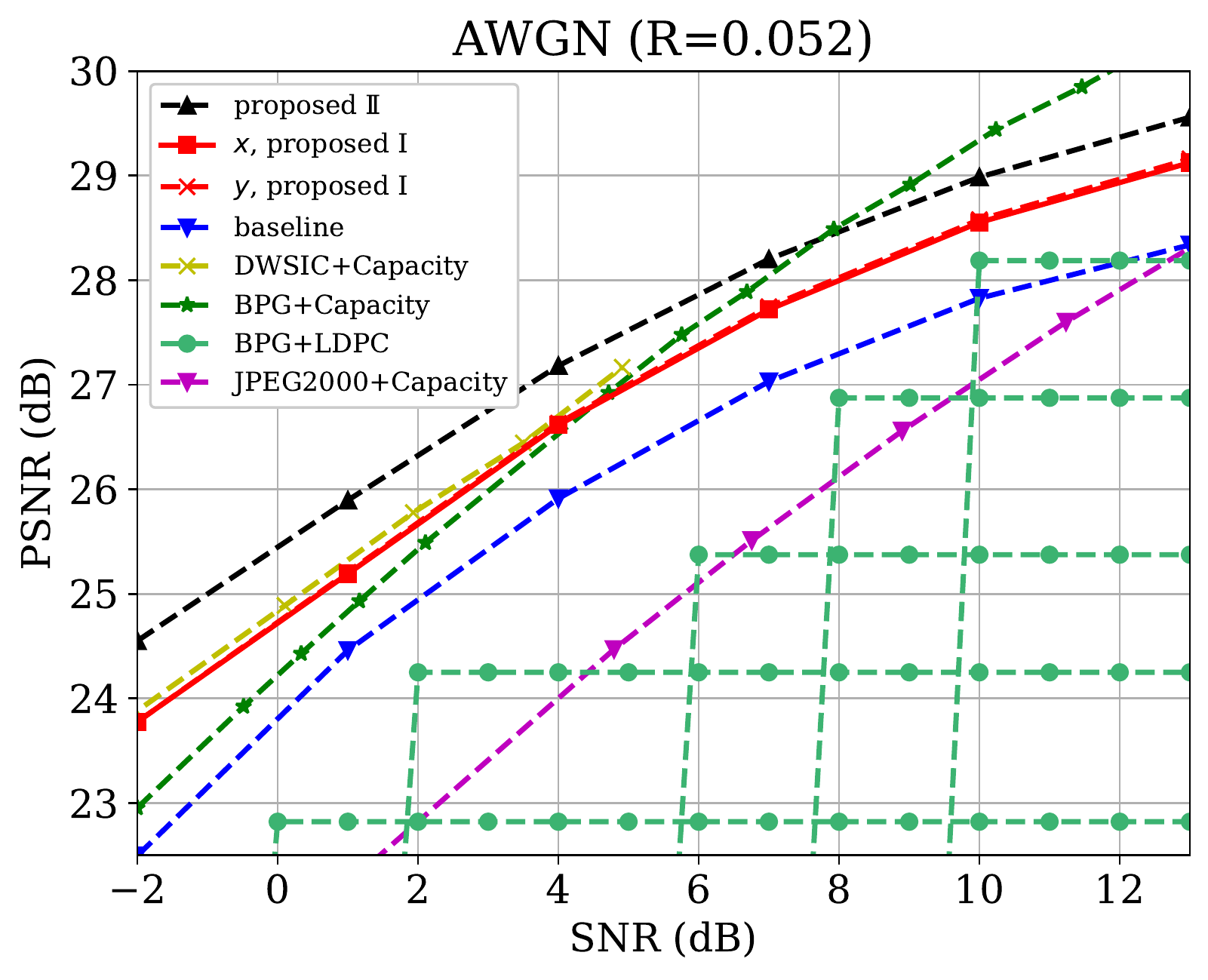}
		}
		\hspace{-.18in}
		\quad
		\subfigure[]{
			
			\includegraphics[width=0.24\textwidth]{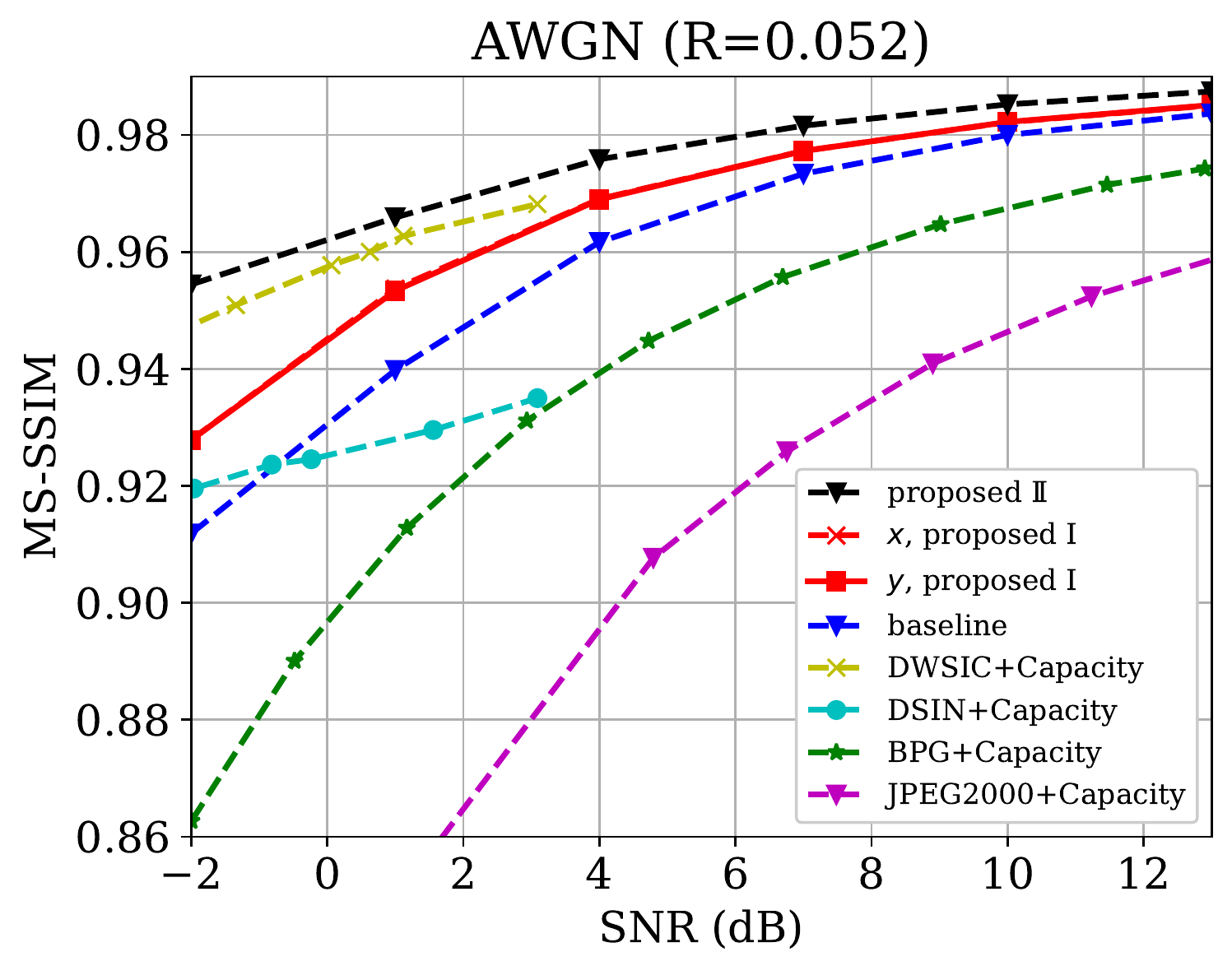}
			
		}
		\hspace{-.18in}
		\quad
		\subfigure[]{
			\includegraphics[width=0.24\textwidth]{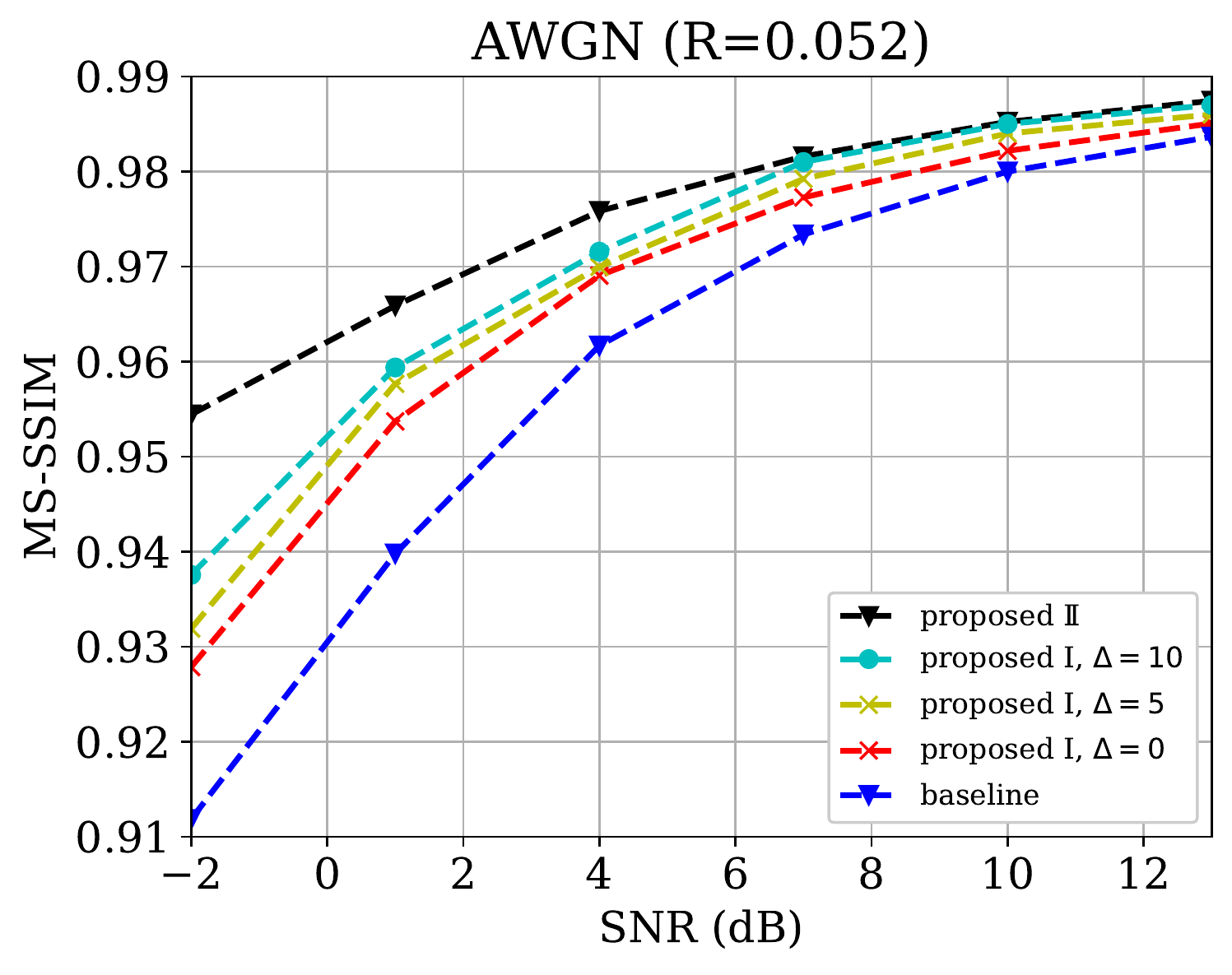}
		}
		\quad
		\hspace{-.18in}
		\subfigure[]{
			\includegraphics[width=0.24\textwidth]{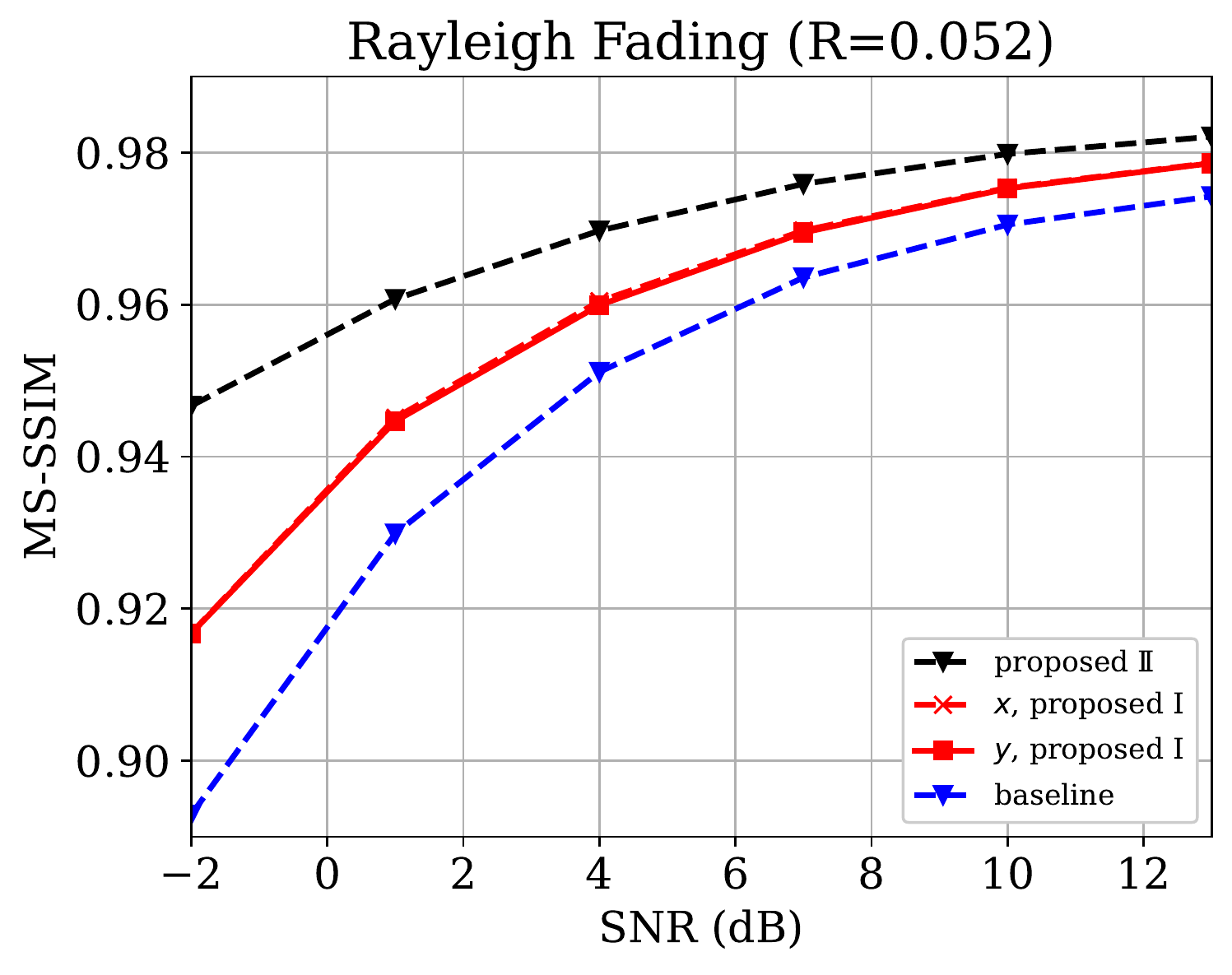}
		}
		\vspace{-0.4cm}
		\caption{The quality metric of reconstructed images under different configurations in different SNRs.}
		\label{fig_curves}
	\end{center}
	\vspace{-0.4cm}
\end{figure*}
\begin{figure*}[htbp]
	\begin{center}
		\subfigure[Original image]{
			\includegraphics[width=0.19\textwidth]{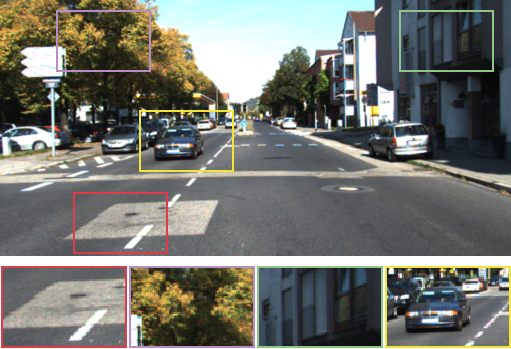}
		}
		\hspace{-.18in}
		\quad
		\subfigure[BPG+Capacity]{
			\includegraphics[width=0.19\textwidth]{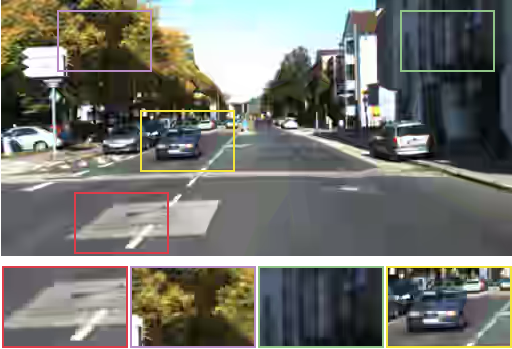}
		}
		\hspace{-.18in}
		\quad
		\subfigure[The baseline scheme]{
			\includegraphics[width=0.19\textwidth]{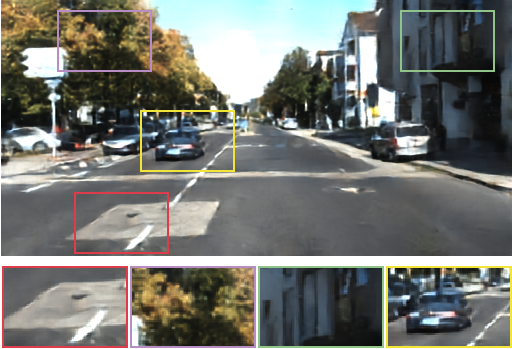}
		}
		\quad
		\hspace{-.18in}
		\subfigure[The proposed scheme I]{
			
			\includegraphics[width=0.19\textwidth]{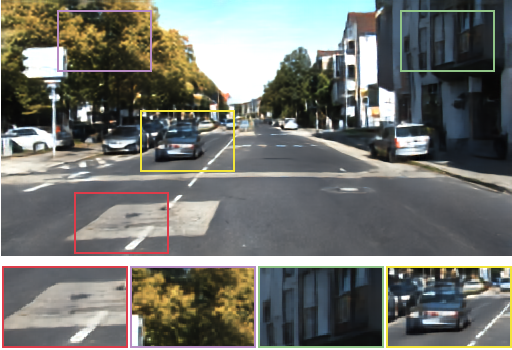}
		}
		\hspace{-.18in}
		\quad
		\subfigure[The proposed scheme II]{
			\includegraphics[width=0.19\textwidth]{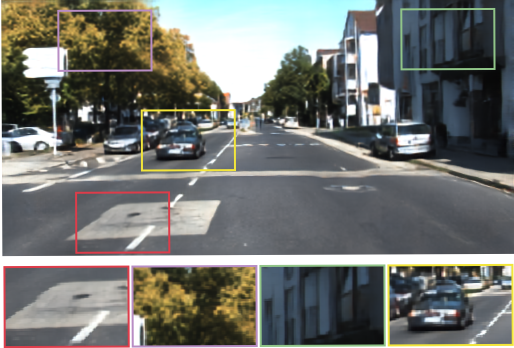}
		}
		\subfigure{\qquad \qquad \qquad  \qquad \qquad \quad }           
		\subfigure{\scriptsize \ MS-SSIM:$0.916$ PSNR:$25.29$dB}
		\subfigure{\scriptsize \ MS-SSIM:$0.938$ PSNR:$23.67$dB}
		\subfigure{\scriptsize \ MS-SSIM:$0.955$ PSNR:$24.76$dB}
		\subfigure{\scriptsize \ MS-SSIM:$0.961$ PSNR:$25.22$dB}
		\caption{Reconstructed images from different source channel coding schemes over $1$dB AWGN channel. The model of proposed scheme is optimized for MS-SSIM.}
		\vspace{-0.4cm}
		\label{fig_reconstruction samples}
	\end{center}
	\vspace{-0.4cm}
\end{figure*}
\subsection{Experimental setup}
\textbf{Datasets.} We constructed our dataset from KITTI Stereo 2012 \cite{KITTI2012} (includes 1578 stereo image pairs) and KITTI Stereo 2015 \cite{KITTI2015} (includes 789 scenes with 21 images pairs per scene taken sequentially). 
Here, a pair of two images means they are taken at the same time from a pair of calibrated stereo cameras.
Following the works on distributed source coding \cite{DSIN,DWSIC}, 1576 image pairs are selected as the training dataset, and 790 image pairs are used for the test. All images are center-cropped and resized to $128\times256$ pixels before training and testing.\\ 
\textbf{Metrics.} Peak signal-to-noise ratio (PSNR) is the most common image evaluation metric, and multi-scale structural similarity (MS-SSIM) is considered to be a perceptual quality metric closer to human perception. Since the two metrics sometimes provide different results, we use both of them to test the robustness of our model.\\
\textbf{Training details.} In all experiments, for the \emph{independent channels case} \cite{DSC3}, the bandwidth ratio of each distributed source $R = 0.052$ (the channel use number per source dimension), which is corresponding to the bottleneck channel dimension $C^{*}=20$. And in the \emph{asymmetric case} \cite{DSC3}, we assume source $\bm{Y}$ is lossless available in the receiver and evaluate the recover quality of $\bm{X}$ in the fixed bandwidth ratio $0.052$. The average transmission power constraint is set as $P=1$, and the two stereo images have the same weight ($\alpha = 1$). The hidden layer size of MLP is set to $4$ times the input dimension. 
Each model is trained for 250K iterations using Adam optimizer with a learning rate of $1\times10^{-4}$ and a batch size of 12. Mean squared error (MSE) is used as the loss function for PSNR models, and $1-\text{MS-SSIM}$ for MS-SSIM models. 
Each model is trained over transmission channel under a uniform distribution of SNR from $-3$dB to $14$dB, and the interval of channel quality tokens is set to $1$dB. Besides, a special token is employed to indicate a noiseless channel for the training of \emph{asymmetric case}.

\subsection{Experimental results}
We compare our scheme with existing state-of-the-art separated design schemes as well as the DL-based JSCC baseline.  
For the source coding of separated design systems, we consider the popular image codecs - BPG and JPEG2000, as well as the recently developed DL-base DSC image compression schemes DSIN \, cite{DSIN}, and DWSIC \cite{DWSIC}. For the channel coding, we employ practical LDPC codes and ideal capacity-achieving code (denoted as Capacity) as bound. 
We denote the combination of source coding and channel coding schemes by "$+$" for brevity. For each configuration of separated systems, given bandwidth ratio, we first calculate the maximum source compression rate $r_{\max}$ (bits per pixel) using channel capacity or the rate of channel code and modulation, then compress the images at the largest rate that no more than $r_{\max}$. 
For LDPC code which cannot guarantee reliable transmission, follow \cite{DJSCCF}, we set the failed reconstruction to the mean value for all the pixels per color channel. The DSIN and DWSIC are digital source coding schemes under lossless side information assumption. To test their transmission performance, we combine a channel code for evaluation. 

As plotted in Fig. \ref{fig_curves}(a) and Fig. \ref{fig_curves}(b), we compare the quality of reconstructed images on the AWGN channel under various channel SNRs in terms of the average MS-SSIM and PSNR. The proposed I refers to the proposed model in \emph{independent channels case} \cite{DSC3} while proposed II refers to the \emph{asymmetric case} \cite{DSC3}. 
The proposed model can adapt to a range of SNRs as well as in the condition when the side information is noiseless, thus all the lines about the proposed model are tested in a single model. 
For the proposed scheme I, we assume both transmission channels have the same SNR and present the results of each metric for two distributed sources. 
Besides, since the proposed JSCD treats each transmission link equally, results show that both distributed sources have almost identical PSNR and MS-SSIM throughout the simulation SNR interval.     

We first compare the proposed method I to the baseline model and classical separated systems. 
As a naive use of D-JSCC, the baseline transmits images $\x$ and $\y$ independently without extra design. 
Due to the proposed scheme leveraging the mutual information of the two links, there is an apparent performance gain compared to the baseline model.  
As for the classical separated systems, our scheme outperforms the classical image codec JPEG2000$+$Capacity and BPG$+$Capacity scheme in terms of MS-SSIM or the low SNR region PSNR. 
Though BPG$+$Capacity shows better PSNR results in the high SNR region, the scheme requires unlimited delay and complexity, which is ideal in the distributed scenario. 
Consider a more practical BPG$+$LDPC scheme. It suffers from the ``cliff effect'' and performs worse than the proposed model in terms of PSNR. 

Then, we also present the results of the proposed method II, which shows the upper bound performance of proposed I. Compared with DL-base DSC schemes DSIN and DWSIC which consider the same case, the proposed scheme shows competitive performance. 
Fig. \ref{fig_curves}(c) presents how the reconstruction quality of one image varies when the channel quality of the other link changes, $\Delta$ denotes the SNR difference of two links. Due to the proposed cross attention module achieving interactions between SNR information and context information of each image, its performance improves with the quality of side information. It adapts to every $\Delta$ value in a single model.
Moreover, the performance of our proposed scheme under a Rayleigh fast fading channel with perfect channel state information is studied in Fig. \ref{fig_curves}(d), which also proves the robustness and the performance gain of the proposed model. A visual example of the reconstructed image over the AWGN channel with $\text{SNR}=1$dB is shown in Fig. \ref{fig_reconstruction samples}. The proposed schemes present a better reconstruction with more details. 

\section{Conclusion}
\label{sec:conclusion}
In this paper, the problem of D-JSCC for correlated image sources has been studied. The proposed JSCE and JSCD structure leverage the common information across two stereo images to improve reconstruction quality without extra transmission overhead. Besides, we propose an SNR-aware cross attention module, which calculates the patch-wise relevance of two images as well as considers the SNR of each transmission link. Due to the channel state information and context information of two images being efficiently exploited, results have demonstrated that the proposed method achieves impressive performance in distributed scenarios.

\vfill
\pagebreak

\bibliographystyle{IEEEbib}
\bibliography{myRef}

\begin{thebibliography}{10}

\bibitem{SW}
D.~Slepian and J.~Wolf,
\newblock ``Noiseless coding of correlated information sources,''
\newblock {\em IEEE Transactions on Information Theory}, vol. 19, no. 4, pp.
  471--480, 1973.

\bibitem{WZ}
A.~Wyner and J.~Ziv,
\newblock ``The rate-distortion function for source coding with side
  information at the decoder,''
\newblock {\em IEEE Transactions on Information Theory}, vol. 22, no. 1, pp.
  1--10, 1976.

\bibitem{DSC1}
Z.~Xiong, A.~D. Liveris, and S.~Cheng,
\newblock ``Distributed source coding for sensor networks,''
\newblock {\em IEEE Signal Processing Magazine}, vol. 21, no. 5, pp. 80--94,
  2004.

\bibitem{DSC2}
S.~S. Pradhan and K.~Ramchandran,
\newblock ``Distributed source coding using syndromes (discus): Design and
  construction,''
\newblock {\em IEEE Transactions on Information Theory}, vol. 49, no. 3, pp.
  626--643, 2003.

\bibitem{DSIN}
S.~Ayzik and S.~Avidan,
\newblock ``Deep image compression using decoder side information,''
\newblock in {\em In European Conference on Computer Vision. (ECCV)}, 2020, pp.
  699--714.

\bibitem{DWSIC}
N.~Mital, E.~Ozyilkan, A.~Garjani, and D.~G{\"u}nd{\"u}z,
\newblock ``Deep stereo image compression with decoder side information using
  wyner common information,''
\newblock {\em arXiv preprint arXiv:2106.11723}, 2021.

\bibitem{xuan2021low}
Z.~Xuan and K.~Narayanan,
\newblock ``Low-delay analog distributed joint source-channel coding using
  sirens,''
\newblock in {\em Proceedings of European Signal Processing Conference
  (EUSIPCO)}. IEEE, 2021, pp. 1601--1605.

\bibitem{DJSCC}
E.~Bourtsoulatze, D.~B. Kurka, and D.~G{\"u}nd{\"u}z,
\newblock ``Deep joint source-channel coding for wireless image transmission,''
\newblock {\em IEEE Transactions on Cognitive Communications and Networking},
  vol. 5, no. 3, pp. 567--579, 2019.

\bibitem{DJSCCF}
D.~B. Kurka and D.~G{\"u}nd{\"u}z,
\newblock ``Deep{JSCC}-f: Deep joint source-channel coding of images with
  feedback,''
\newblock {\em IEEE Journal on Selected Areas in Information Theory}, vol. 1,
  no. 1, pp. 178--193, 2020.

\bibitem{ADJSCC}
J.~Xu, B.~Ai, W.~Chen, A.~Yang, P.~Sun, and M.~Rodrigues,
\newblock ``Wireless image transmission using deep source channel coding with
  attention modules,''
\newblock {\em IEEE Transactions on Circuits and Systems for Video Technology},
  early access, 2021.

\bibitem{wang2021novel}
S.~Wang, J.~Dai, S.~Yao, K.~Niu, and P.~Zhang,
\newblock ``A novel deep learning architecture for wireless image
  transmission,''
\newblock in {\em Proceedings of IEEE Global Communications Conference}, 2021.

\bibitem{DSC3}
J.~Garcia-Frias and Z.~Xiong,
\newblock ``Distributed source and joint source-channel coding: from theory to
  practice,''
\newblock in {\em Proceedings of IEEE International Conference on Acoustics,
  Speech and Signal Processing. (ICASSP)}, 2005, vol.~5, pp. v/1093--v/1096
  Vol. 5.

\bibitem{HIFIC}
F.~Mentzer, G.~Toderici, Michael M.~Tschannen, and E.~Agustsson,
\newblock ``High-fidelity generative image compression,''
\newblock {\em in Advances in Neural Information Processing Systems}, vol. 33,
  2020.

\bibitem{VIT}
A.~Dosovitskiy, L.~Beyer, A.~Kolesnikov, D.~Weissenborn, X.~Zhai,
  T.~Unterthiner, M.~Dehghani, M.~Minderer, G.~Heigold, S.~Gelly, et~al.,
\newblock ``An image is worth 16x16 words: Transformers for image recognition
  at scale,''
\newblock {\em Proceedings of International Conference on Learning
  Representations. (ICLR)}, 2020.

\bibitem{KITTI2012}
A.~Geiger, P.~Lenz, and R.~Urtasun,
\newblock ``Are we ready for autonomous driving? the kitti vision benchmark
  suite,''
\newblock pp. 3354--3361, 2012.

\bibitem{KITTI2015}
M.~Menze, C.~Heipke, and A.~Geiger,
\newblock ``Joint 3d estimation of vehicles and scene flow,''
\newblock {\em ISPRS Annals of the Photogrammetry, Remote Sensing and Spatial
  Information Sciences}, vol. 2, pp. 427, 2015.

\end{thebibliography}

\end{document}